\documentclass[a4paper,11pt]{article}
\usepackage{pos}
\usepackage{amsmath}

\newcommand{\be}{\begin{equation}}\newcommand{\ee}{\end{equation}}
\newcommand{\bea}{\begin{eqnarray}}\newcommand{\eea}{\end{eqnarray}}
\newcommand{\brr}{\begin{array}}\newcommand{\err}{\end{array}}
\newcommand{\bit}{\begin{itemize}}\newcommand{\eit}{\end{itemize}}
\newcommand{\ben}{\begin{enumerate}}\newcommand{\een}{\end{enumerate}}

\newcommand{\bbm}{\begin{bmatrix}}\newcommand{\ebm}{\end{bmatrix}}

\newcommand{\ba}{\begin{array}}
\newcommand{\ea}{\end{array}}

\def\lf{\left}

\def\ran{\rangle}

\def\ri{\right}

\def\de{\delta}

\def\si{\sigma}\def\Si{\Sigma}

\def\1{{_{1}}}\def\2{{_{2}}}

\def\noHe0{:\;\!\!\;\!\!:H_e(0):\;\!\!\;\!\!:}
\def\noHm0{:\;\!\!\;\!\!:H_\mu(0):\;\!\!\;\!\!:}

\def\lf{\left}

\def\ran{\rangle}

\def\ri{\right}

\def\de{\delta}

\def\si{\sigma}\def\Si{\Sigma}

\def\1{{_{1}}}\def\2{{_{2}}}

\def\I{{_{\rm{I}}}}\def\II{{_{\rm{II}}}}

\title{Bekenstein bound from the Pauli principle: a brief introduction}

\author[a]{Giovanni Acquaviva}
\author[b]{Alfredo Iorio}
\author*[b]{Luca Smaldone}

\affiliation[a]{Institute of Theoretical Physics, Faculty of Mathematics and Physics, Charles University\\ V Holesovi\v{c}k\'{a}ch 2, 18000 Praha 8, Czech Republic}

\affiliation[b]{Institute of Particle and Nuclear Physics, Faculty of Mathematics and Physics, Charles University\\ V Holesovi\v{c}k\'{a}ch 2, 18000 Praha 8, Czech Republic}

\emailAdd{gioacqua@utf.troja.mff.cuni.cz}
\emailAdd{iorio@ipnp.troja.mff.cuni.cz}
\emailAdd{smaldone@ipnp.mff.cuni.cz}

\abstract{Here we briefly resume the idea, originally introduced in Phys. Rev. D 102, 106002 (2020), that the Bekenstein bound on entropy is a consequence of the fermionic nature of fundamental degrees of freedom, which arrange themselves to form matter and spacetime. The main point is discussed by means of a toy-model of black hole evaporation, which describes the dynamics of such degrees of freedom, called $X$ons. An intrinsic notion of interior/exterior of the black hole during the evaporation process is given and both von Neumann and black hole/environment entropies are computed.}

\FullConference{%
  40th International Conference on High Energy physics - ICHEP2020\\
  July 28 - August 6, 2020\\
  Prague, Czech Republic (virtual meeting)
}

\begin{document}
\maketitle

The Bekenstein Bound \cite{beke1981} states that a bounded physical system at fixed energy can store a finite amount of entropy; such bound is exactly saturated for a black hole (BH), in which case its entropy coincides with the Bekenstein-Hawking one.  In \cite{acqua2017} two main consequences of this bound have been presented: i) the degrees of freedom (DoF) responsible for the BH entropy have to take into account both matter and spacetime and hence must be of a new, more fundamental nature than the DoF we know: here we call such DoF “$X$ons”; and ii) the Hilbert space of the $X$ons of a given BH is necessarily finite dimensional \cite{bao2017}.  Here we adopt the opposite point of view \cite{acqua2020}: by focusing on the $X$ons level, where there are no separate notions of fields and geometry, we assume that in a BH only free $X$ons exist, and that they are finite in number and fermionic in nature.  Hence any quantum level can be filled by no more than one fermion: this assures that the Hilbert space $\mathcal{H}$ of physical states with a finite number of levels is finite dimensional.  Given the total number $N$ of quantum levels available to the BH, the evaporation is the process $N \to (N -1) \to (N - 2) \to \cdots$ in which the number of \textit{free} $X$ons steadily decreases in favor of the $X$ons that are \textit{rearranged} into quasi-particles and the spacetime they propagate on.  At the $X$ons level: i) there is no pre-existing \textit{time}, because the natural evolution parameter is the average number of free $X$ons; and ii) there is no pre-existing \textit{space}.  However, the steady process of evaporation described above gives us a natural distinction between the \textit{outside} (region $\mathrm{I}$) and the \textit{inside} (region $\mathrm{II}$) of the BH.  The Hilbert space of \textit{physical states} $\mathcal{H}$ is built as a subspace of a larger \textit{kinematical} Hilbert space $\mathcal{K} \ \equiv \ \mathcal{H}_\I \otimes \mathcal{H}_\II$, and it has dimension $\Si \ \equiv \ {\rm dim} \, \mathcal{H} \ = \ 2^N$.  We introduce the fermionic ladder operators
\be
\lf\{a_{n} \, , \, a^\dag_{n'}\ri\} \ = \ \lf\{b_{n} \, , \, b^\dag_{n'}\ri\} \ = \  \de_{n n^{'}} \, , 
\ee
with $n,n'= 1, \dots, N$ and $a$, $b$ represents \emph{environment} and \emph{black hole} modes, respectively. All other anticommutators equal to zero.  We initialize the system in a pure state representing the BH at the beginning of the evaporation process, with all the slots occupied by free $X$ons. In the final state the $X$ons are all recombined to form fields and spacetime.  A toy-model corresponding to such boundary conditions is defined by the entangled state
\be \label{alexp}
|\Psi(\si)\ran \ = \ \prod^N_{i=1} \, \sum_{n_i=0,1} \, C_i(\si) \, \lf(a^{\dag}_i\ri)^{n_i} \, \lf(b^{\dag}_i \ri)^{1-n_i}  \, |0\ran_{\I} \otimes |0\ran_{\II} \, ,
\ee
with $C_i= (\sin \si)^{n_i} \, (\cos \si)^{1-n_i}$. The parameter $\si$ can be seen as an interpolating parameter, which describes the evolution of the system from $\si = 0$ until $\si = \pi/2$, respectively the beginning and the end of the evaporation.  Following the framework of Thermofield dynamics (TFD) \cite{ume1996}, we quantify entanglement for environment and BH modes at each stage through the entropy operators
\bea
S_{\I}(\si) \ &= \ -\sum^N_{n=1} \, \lf(a^\dag_n \, a_n \, \ln \sin^2 \si + a_n \, a^\dag_n \, \ln \cos^2 \si\ri) \, .\\
S_{\II}(\si) \ &= \ -\sum^N_{n=1} \, \lf(b^\dag_n \, b_n \, \ln \cos^2 \si + b_n \, b^\dag_n \, \ln \sin^2 \si\ri) \, .
\eea
As it must be for bipartite systems, the averages of the two operators coincide:
\be
\mathcal{S}_\I (\si) \ = \ \mathcal{S}_\II  (\si) \ = \  -N \lf(\sin^2 \si \, \ln \sin^2 \si + \cos^2\si \, \ln \cos^2 \si\ri) \, .
\ee
Such average is the von Neumann entropy quantifying the entanglement between environment and BH: it starts and ends in zero, while its maximum value is $\mathcal{S}_{max} \ = \ N \, \ln 2 \ = \ \ln \Si$, so that $\Si \ = \  e^{\mathcal{S}_{max}}$.  Hence, in our model $\mathrm{dim} \,\mathcal{H} $ is related to the maximal entanglement entropy of the environment with the BH. This happens exactly when the modes have \textit{half} probability to be inside and \textit{half} probability to be outside the BH, and then the largest amount of bits are necessary to describe the system. The system has thus an intrinsic way to know how big is the physical Hilbert space, which is related to the initial size of the BH: the maximal entanglement $\mathcal{S}_{max}$ tells how big was the original BH. Hence $\mathcal{S}_{max}$ must be some function of the original mass ${\cal M}_0$ of the BH. This is the Bekenstein bound in this picture, obtained as a consequence of a Pauli principle at $X$ons level. It is important to stress that only a subspace $\mathcal{H} \subseteq  \mathcal{H}_\I \otimes  \mathcal{H}_\II$ (with ${\rm dim} \, \mathcal{H}= 2^N$) of the full Hilbert space is the one of physical states.  Nonetheless, one could think that the physical Hilbert spaces of the two subsystems should take into account only the number of modes truly occupied at any given stage of the evaporation. Hence, the actual dimensions would be $2^{N_\I (\si)}$ and $2^{N_{\II} (\si)}$, where the mean numbers of modes on $|\Psi(\si)\ran$ are
\be
N_\I (\si) \ \ = \ N \sin^2 \si \, , \qquad  N_\II (\si) \ = \  N- N_\I(\si) \ = \ N \, \cos^2 \si \, . 
\ee
It becomes clear that $\sigma$ is a discrete parameter in the interval $[0, \pi/2]$, essentially counting the diminishing number of free $X$ons. When we take this view it is natural to consider the partition $2^N = 2^{N_{\II}(\si)} \times 2^{N_\I(\si)} \equiv n \times m$, with $n = 2^N, 2^{N - 1}, \dots, 1$, and $m = 1, \dots, 2^{N - 1}, 2^N$. 
Hence we define the Bekenstein and environment entropy as
\bea
{\cal S}_{BH} \ \equiv \ \ln n = N \ln 2 \cos^2 \sigma \,, \qquad {\cal S}_{env} \ \equiv \ \ln m = N \ln 2 \sin^2 \sigma \,,
\eea
which satisfy the inequality $\mathcal{S}_\I \ \leq \ \mathcal{S}_{BH} \, + \, \mathcal{S}_{env} \ = \  \mathcal{S}_{max}$. In the figure, we plot ${\cal S}_{BH}$ (black), ${\cal S}_{env}$ (green) and ${\cal S}_{I}$ (red) for $N=1000$.
\begin{center}
\includegraphics[width=0.35\textwidth]{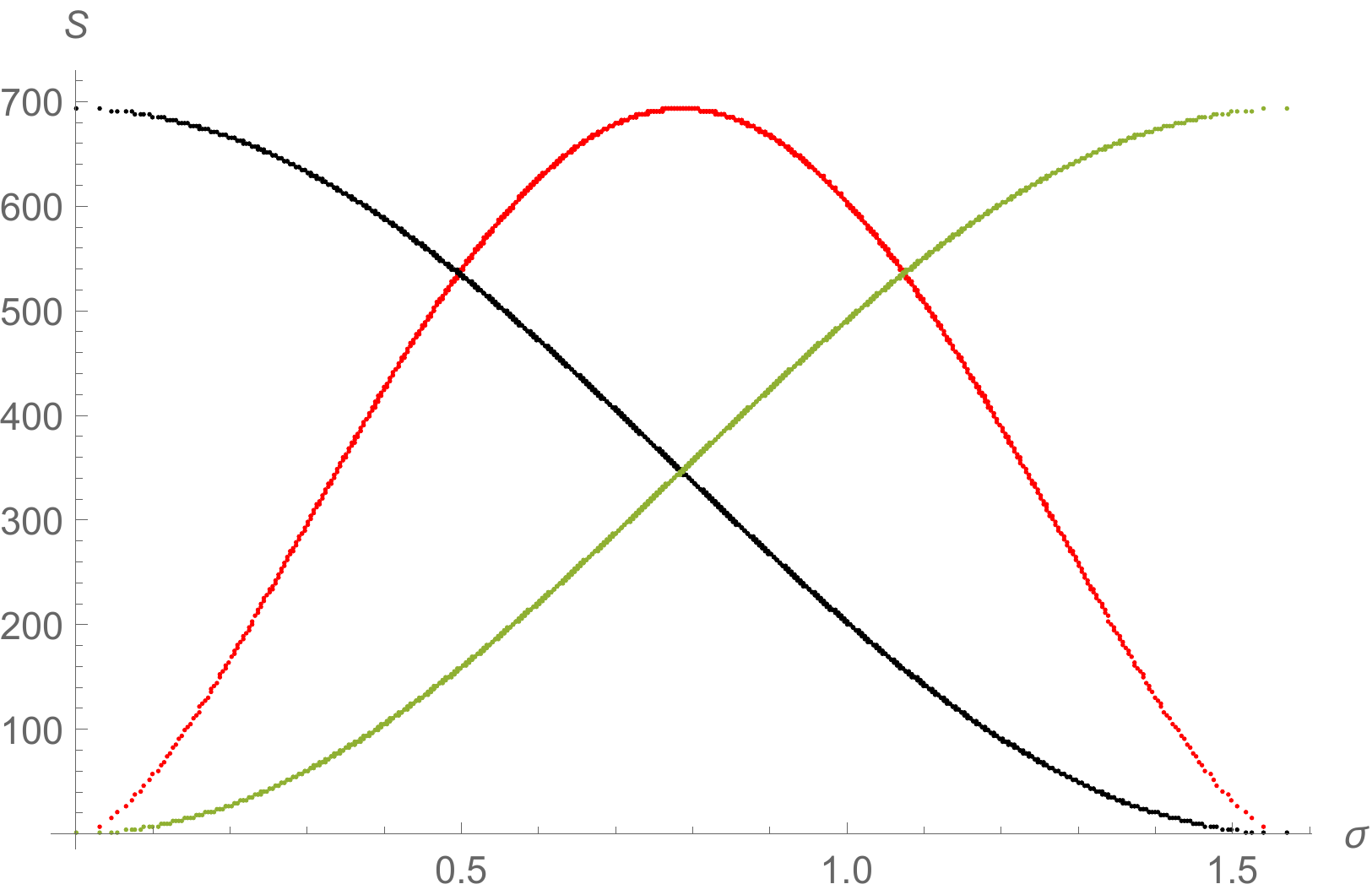}
\end{center}
For a full identification of ${\cal S}_{max}$ with the entropy ${\cal S}_{BH}$ of the initial BH, one would have (for a non-rotating, uncharged black hole) $N \ = \ \frac{4 \, \pi \, \mathcal{M}^2_0}{l_P^2\, \ln 2}$, where $l_P$ is the Planck length.

 A.I. and L.S. acknowledge support from Charles University Research Center (UNCE/SCI/013).


\begin{thebibliography}{99}
\bibitem{beke1981} J. D. Bekenstein, Phys. Rev. D 23, 287 (1981).
\bibitem{acqua2017}  G. Acquaviva, A. Iorio and M. Scholtz, Ann. Phys. 387, 317 (2017).
\bibitem{bao2017} N. Bao, S. M. Carroll and A. Singh, Int. J. Mod. Phys. D 26, no. 12, 1743013 (2017).
\bibitem{acqua2020} G. Acquaviva, A. Iorio and L. Smaldone, Phys. Rev. D 102, 106002 (2020).
\bibitem{ume1996} Y. Takahashi and H. Umezawa, Int. J. Mod. Phys. B 10, 1755 (1996).
\end{thebibliography}
\end{document}